\documentclass[12pt,preprint]{aastex} 
\usepackage{multirow}

\newcommand{\kms}{km~s$^{-1}$} 
\newcommand{\mum}{$\mu$m}

\newcommand{\methanol}{CH$_3$OH}
\newcommand{\ammonia}{NH$_3$}

\newcommand{\mjb}{mJy~beam$^{-1}$}
\newcommand{\HII}{H\,{\sc ii}}

\newcommand{\cml}{centimeter}
\newcommand{\cmc}{centimeter continuum}
\newcommand{\cc}{cm$^{-3}$}
\newcommand{\ct}{cm$^{-2}$}

\begin{document}

\shortauthors{Brogan et al.}

\shorttitle{1.3~cm EVLA Survey of G35.03+0.35 } 

\title{First Results from a 1.3~cm EVLA Survey of \\ Massive Protostellar
  Objects: G35.03+0.35 }

\author{C. L. Brogan\altaffilmark{1}, T. R. Hunter\altaffilmark{1},
  C. J. Cyganowski\altaffilmark{2,3},\\ R. K. Friesen\altaffilmark{1},
  C. J. Chandler\altaffilmark{5}, R. Indebetouw\altaffilmark{1,6}  }
 
\email{cbrogan@nrao.edu}

\altaffiltext{1}{NRAO, 520 Edgemont Rd, Charlottesville, VA 22903} 
\altaffiltext{2}{Harvard-Smithsonian Center for Astrophysics,
  Cambridge, MA 02138}
 \altaffiltext{3}{NSF Astronomy and Astrophysics Postdoctoral Fellow} 
\altaffiltext{5}{NRAO, 1003 Lopezville Rd, Socorro, NM 87801} 
\altaffiltext{6}{University of Virginia, Charlottesville, VA 22903}  

\slugcomment{\footnotesize to appear in {\it The Astrophysical Journal Letters} EVLA Special Issue}

\begin{abstract}

We have performed a 1.3~cm survey of 24 massive young stellar objects
(MYSOs) using the Expanded Very Large Array (EVLA).  The sources in
the sample exhibit a broad range of massive star formation signposts
including Infrared Dark Clouds (IRDCs), UCHII regions, and extended
4.5\mum\ emission in the form of Extended Green Objects (EGOs).  In
this work, we present results for G35.03+0.35 which exhibits
all of these phenomena.  We simultaneously image the 1.3~cm
\ammonia\ (1,1) through (6,6) inversion lines, four
\methanol\ transitions, two H recombination lines, plus continuum at
0.05~pc resolution.  We find three areas of thermal
\ammonia\ emission, two within the EGO (designated the NE and SW
cores) and one toward an adjacent IRDC.  The NE core contains an
UC\HII\ region (CM1) and a candidate HC\HII\ region (CM2).  A region
of non-thermal, likely masing \ammonia\/ (3,3) and (6,6) emission is
coincident with an arc of 44 GHz \methanol\ masers.  We
also detect two new 25~GHz Class I \methanol\/ masers. A complementary
Submillimeter Array 1.3~mm continuum image shows that the distribution
of dust emission is similar to the lower-lying \ammonia\/ lines, all
peaking to the NW of CM2, indicating the likely presence of an
additional MYSO in this protocluster.  By modeling the \ammonia\/ and
1.3~mm continuum data, we obtain gas temperatures of 20-220 K and
masses of 20-130 M$_{\sun}$. The diversity of continuum emission
properties and gas temperatures suggest that objects in a range of
evolutionary states exist concurrently in this protocluster.

\end{abstract}
 
\keywords{stars: formation --- stars: massive --- 
ISM: individual objects (G35.03+0.35) ---  techniques:
interferometric
}

\section{Introduction}

Massive star formation is a phenomenon of fundamental importance in
astrophysics, yet it remains poorly understood.  Massive protostars
form in complex clusters and predominately at distances greater than a
kiloparsec making them challenging to study \citep{Zinnecker07}.
Unfortunately, the tools used to study low mass YSOs (primarily
near-IR imaging) are largely inapplicable due to extreme dust
obscuration even into the mid-IR.  Radio and millimeter wavelengths
penetrate the dust, and have revealed a wide variety of phenomena
associated with MYSOs: \methanol\/, H$_2$O, and OH masers,
hyper-compact (HC) and ultra-compact (UC) \HII\/ regions,
recombination lines, infrared dark clouds (IRDCs), warm ($> 30$~K)
dust cores, massive outflows, extended 4.5 \mum\/ emission from
shocks, and hot core line emission.  However, these signposts have
been compiled from a heterogeneous set of observations with varying
angular resolution and sensitivity, making correlation analyses
difficult.

In order to advance our understanding of MYSOs, we are using the EVLA
to observe 24 MYSOs with $\sim 10,000$ AU resolution in 1.3~cm
continuum and a comprehensive set of diagnostic lines {\em
  simultaneously}. This resident shared risk observing project is
among the first to utilize 16 spectral windows with high spectral
resolution including: the \ammonia\/ ladder from (1,1) to (6,6)
($E_l$=23 - 540 K); four \methanol\/ transitions that can show maser
emission; two H recombination lines; and other potential hot core
species; plus reasonable continuum sensitivity.  The majority (22/24)
of our sample originates from the \citet{Cyganowski08} catalog of MYSO
outflow candidates (extended green objects, EGOs, selected based on
extended 4.5\mum\ emission). Many EGOs are located in IRDCs \citep[for
  an overview of IRDCs see][]{Rathborne06}, with a smaller subset
associated with HC or UCHII regions. To explore the range in the
properties of MYSOs associated with extended 4.5\mum\/ shock emission,
the sample includes five EGOs with compact HII regions, as well as two
non-EGOs for comparison (an IRDC and a UCHII region).  In this paper,
we present the first results from the survey by describing the
observations of one source in detail: G35.03$+$0.35 (hereafter,
G35.03).  To show the location of dust emission with respect to the
centimeter emission, we include complementary 1.3~mm Submillimeter
Array (SMA) observations.

\subsection{Background on G35.03$+$0.35}

G35.03 is an EGO at a distance of $\sim 3.4$ kpc with a bipolar 4.5
\mum\/ morphology oriented NE-SW at a systemic velocity of $\sim 53.1$
\kms\/ based on H$^{13}$CO$^{+}$ (3-2) single dish data
\citep{Cyganowski09}.  A saturated {\em Spitzer} MIPSGAL 24~\mum\/
source is located near the center of the two lobes and is coincident
with five compact \cmc\/ sources called CM1...CM5 in order of
decreasing flux \citep[][]{Cyganowski11}. CM1 has a flat \cml\/
spectral index ($\alpha\sim -0.1$, $S_{\nu}\propto \nu^{\alpha}$)
consistent with optically thin free-free emission from an UC \HII\/
region \citep[also see][]{Kurtz94,Argon00}. In contrast CM2 has a
rising spectrum ($\alpha\sim +0.7$), suggesting it may be a HC~\HII\/
region or wind/jet source \citep{Cyganowski11}.  Coincident with CM2
are blueshifted Class-II 6.7~GHz \methanol\/ masers \citep[$\Delta
  V\sim 6-12$ \kms\/;][]{Cyganowski09}, blueshifted OH masers
\citep[$\Delta V\sim 3-14$ \kms\/;][]{Argon00}, and redshifted H$_2$O
masers \citep[$\Delta V\sim 13-17$ \kms\/;][]{Forster99}.
Additionally, an arc-like structure of outflow-tracing Class-I 44 GHz
\methanol\/ masers are located toward the Eastern side of the EGO
\citep{Cyganowski09}.

\section{Observations}

The NRAO\footnote{The National Radio Astronomy Observatory is a
  facility of the National Science Foundation operated under agreement
  by the Associated Universities, Inc.}  Expanded Very Large Array
(EVLA) observations are summarized in Table~\ref{smaobs}; see
\citet{Perley11} for additional details on the EVLA system.  The EVLA
data were calibrated and imaged in
CASA\footnote{http://casa.nrao.edu/}.  All images were restored with a
synthesized beam of $3\farcs7\times 3\farcs$0 (P.A.=$-50\arcdeg$),
corresponding to $\sim 12,600\times 10,000$ AU at 3.4~kpc.

The SMA\footnote{The Submillimeter Array (SMA) is a collaborative
  project between the Smithsonian Astrophysical Observatory and the
  Academia Sinica Institute of Astronomy \& Astrophysics of Taiwan.}
observations are also summarized in Table~1.  The sidebands were
centered at 220.1 and 230.1 GHz. The data were calibrated in MIRIAD,
then exported to CASA for self-calibration and imaging. In this paper
we only consider the 1.3~mm continuum data estimated from line-free
channels in the uv-plane. The SMA spectral line data will be described
in a future publication.

\section{Results} 

\begin{figure}
\epsscale{1.0}
\includegraphics[height=3.5in]{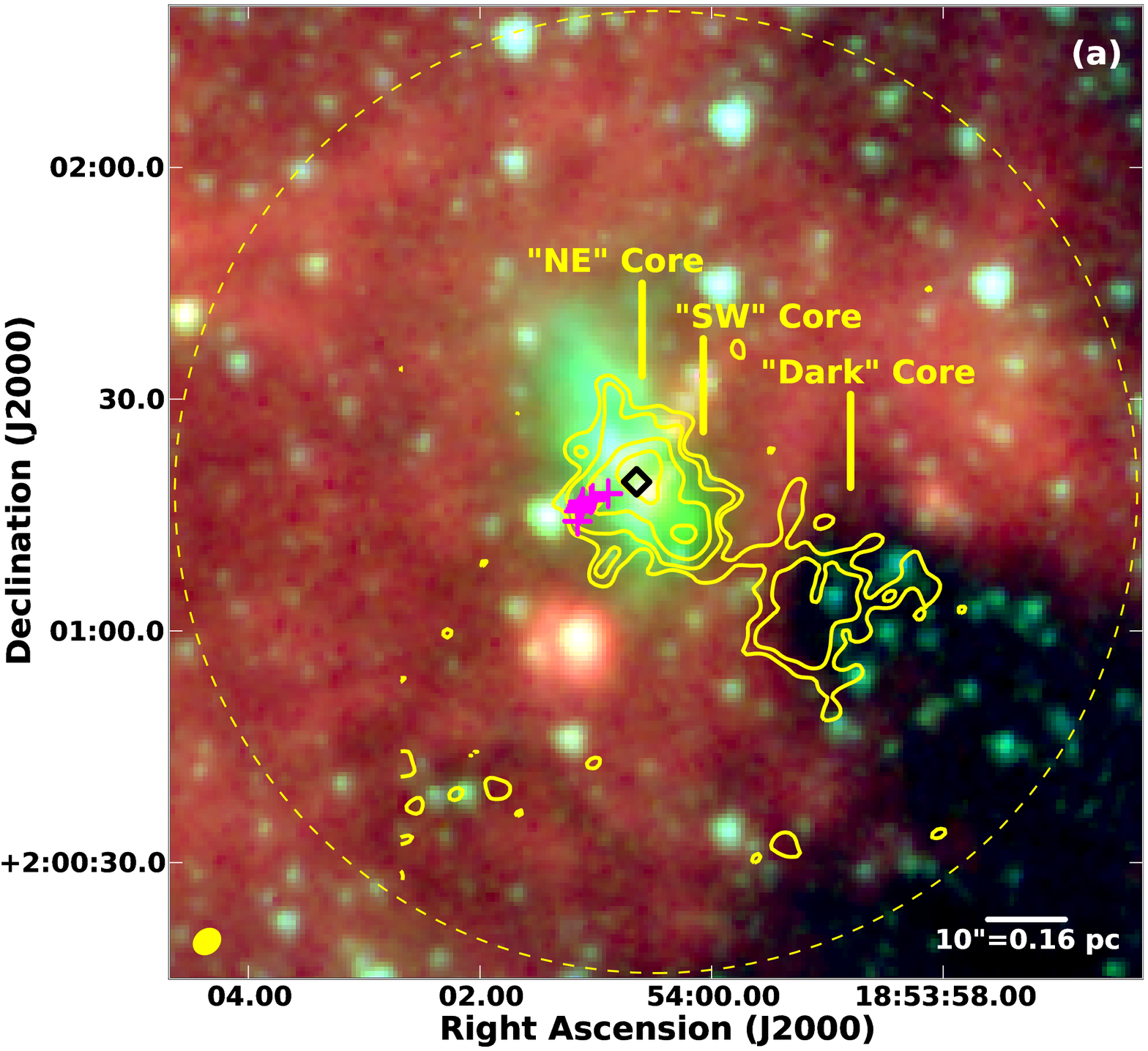}
\includegraphics[height=3.5in]{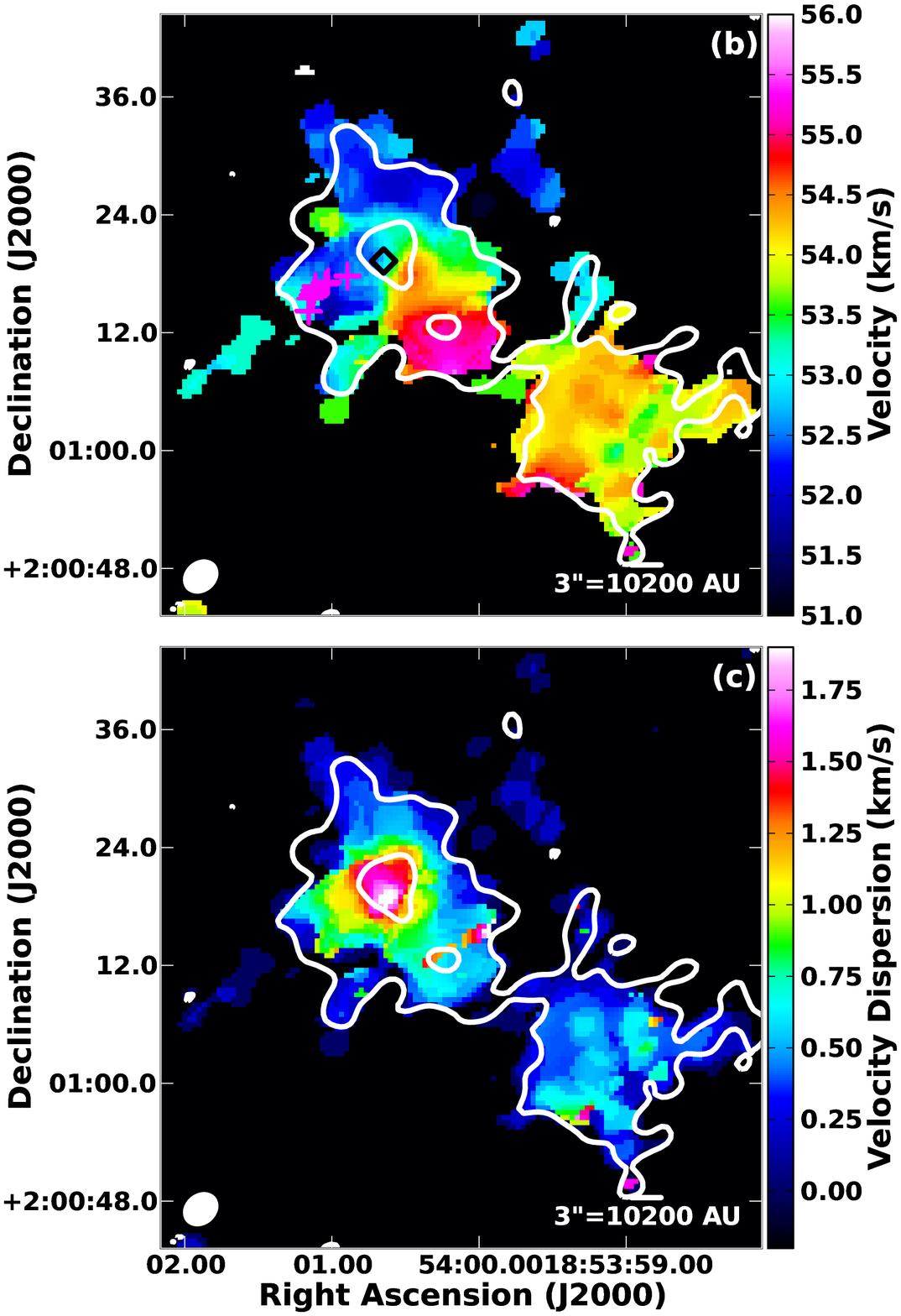}
\caption{(a) {\em Spitzer} GLIMPSE three-color image of EGO G35.03 
  with RGB mapped
  to 8.0, 4.5, and 3.6 \mum\/ overlaid with the EVLA primary beam
  (yellow dashed circle). Yellow contours show
  the \ammonia\/ (1,1) integrated intensity (contours at 16, 28,
  56, and 102 \mjb\/*\kms\/). The three brightest \ammonia\/ cores are
  labeled for reference. Magenta $+$ symbols and a
  single black $\diamond$ mark the locations of the 44 and 6.7 GHz
  \methanol\/ masers from \citet{Cyganowski09}.  (b) and (c) show the
  \ammonia\/ (1,1) velocity field (moment 1) and velocity dispersion
  (moment 2) with the lowest and highest (1,1) integrated intensity
  contour levels from (a) superposed in white. In all three panels, the
  EVLA synthesized beam is shown in the lower left.
 \label{fig1}}
\end{figure}

As shown in Figure~\ref{fig1}a, we detect three major regions of
\ammonia\/ (1,1) emission: (i) the strongest emission is coincident
with the central region of the bipolar 4.5 \mum\/ nebulosity, the five
\cml\/ continuum sources described in \S1.1, and 6.7~GHz \methanol\/
masers; (ii) a slightly weaker and more compact core coincident with
the SW lobe of the bipolar 4.5 \mum\/ nebulosity; and (iii) a weak but
extended core coincident with the IRDC $\sim 30\arcsec$ SW of the EGO.
Figures~\ref{fig1}b and c show the intensity weighted velocity field
and velocity dispersion of the \ammonia\/ (1,1) emission,
respectively.  The emission spans a velocity range of about 4 \kms\/,
with a distinct NE-SW blue-red velocity gradient across the EGO. This
velocity gradient is not clearly associated with an outflow, but
instead seems to trace a shift in systemic velocity across the region.
The mean systemic velocity toward the EGO and the ``Dark'' core to the
SW is $\sim 54$ \kms\/.  In the vicinity of the ``NE'' core (and the
\cml\/ sources), the velocity dispersion shows a marked increase from
$\sim 0.6$ \kms\/ to values as high as $\sim 1.9$ \kms\/. There are also
''fingers'' of weak \ammonia\/ emission evident around the periphery of
the EGO -- these fingers are not apparent in the integrated intensity
map (Fig.~\ref{fig1}a) because as demonstrated in Fig.~\ref{fig1}c
they have extremely narrow velocity dispersion ($\sim 0.25$ \kms\/).

\begin{figure}
\epsscale{1.0}
\includegraphics[height=5.5in]{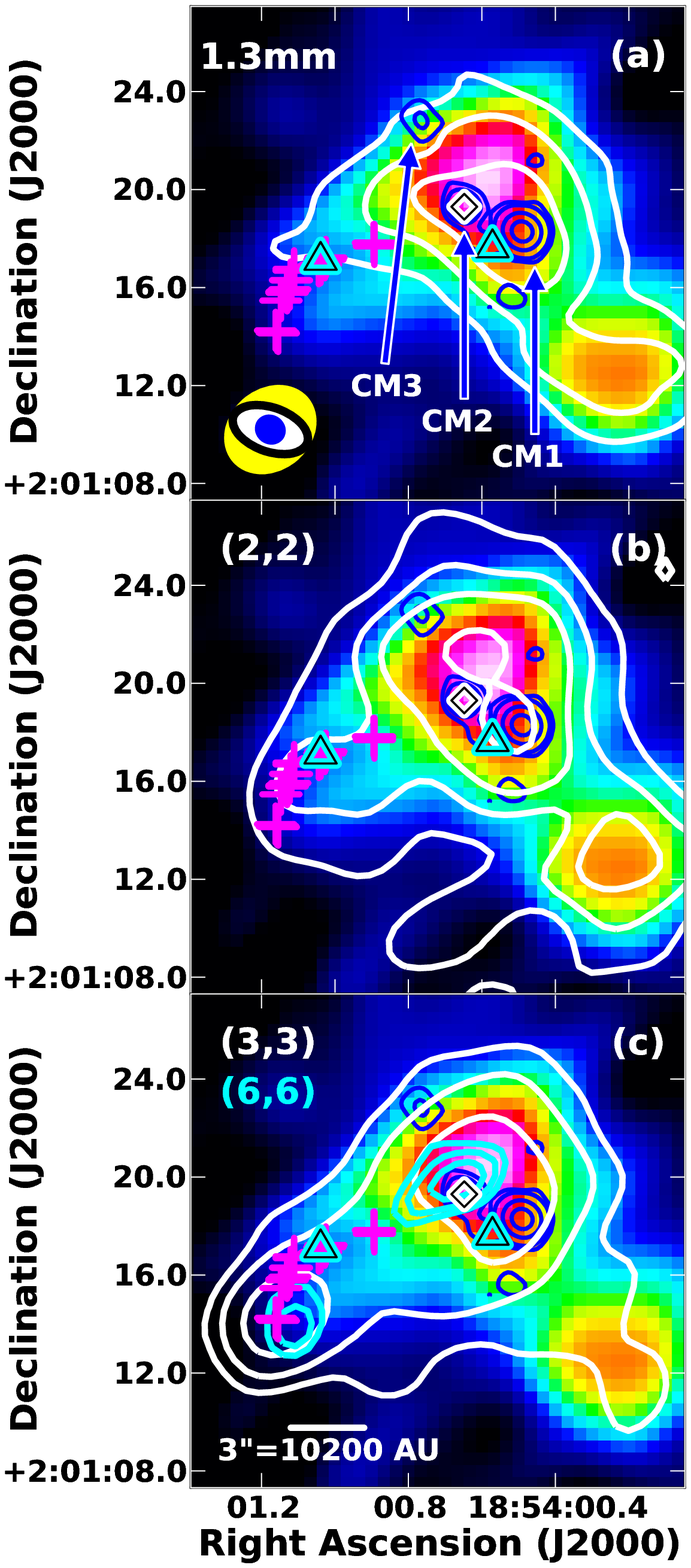}
\caption{Zoomed-in view of the NE and SW cores showing the
  \ammonia\ (1,1) integrated intensity in colorscale with VLA 3.6~cm
  continuum contours (0.1, 0.3, 2, 7 \mjb\/) from \citet{Cyganowski11}
  superposed in blue; the three brightest \cmc\ sources are
  labeled CM1, CM2, and CM3. (a) SMA 1.3~mm continuum is shown in
  white contours (levels: 7, 30, and 90 \mjb\/) and the synthesized
  beams of the EVLA (yellow), VLA (blue), and SMA (white) are shown in
  the lower left corner. (b) White contours show the integrated
  intensity of \ammonia\/ (2,2) (levels: 25, 50, 75, and 117
  \mjb\/*\kms\/).  (c) Contours show the integrated intensity of
  \ammonia\ (3,3) (white; 30, 60, and 120 \mjb\/*\kms\/) and (6,6)
  (cyan; 13, 17, and 22 \mjb\/*\kms\/). In all panels, the locations
  of two new 25 GHz \methanol\/ masers are cyan triangles, and the 6.7 and
  44 GHz \methanol\/ masers from Fig~\ref{fig1} are also shown.
 \label{fig2}}
\end{figure}

Figure~\ref{fig2}a shows a detailed comparison of the EVLA \ammonia\/
(1,1) integrated intensity, SMA 1.3~mm continuum, and the VLA 3.6~cm
continuum \citep[][CM1 and CM2 are also detected in the line free
  channels of the current EVLA data; albeit at lower angular
  resolution and sensitivity]{Cyganowski11}. Overall, there is strong
agreement between the \ammonia\/ and 1.3~mm continuum emission. The
strongest 1.3~mm emission arises from the NE core, coincident with CM1
and CM2. Though it is difficult to quantify the free-free vs. dust
contributions at 1.3~mm due to the mismatch in the continuum
resolution (and $uv-$coverage) between the SMA, VLA, and EVLA (not
shown), it is clear that a significant fraction must be due to
dust. The morphology of the \ammonia\/ (1,1) and (2,2) emission (see
Fig.~\ref{fig2}b), along with the 1.3~mm emission to the NW of CM2,
suggests the presence of a third compact source in the NE core lacking
in free-free emission. Millimeter continuum emission is also visible
to the East-SE of the NE-core that partially coincides with the arc of
44~GHz \methanol\/ masers, and mimics the morphology of the \ammonia\/
(1,1) integrated intensity.  The SW \ammonia\/ core also has a 1.3~mm
counterpart. The ``Dark'' ammonia core was not detected by the SMA,
probably due to the smaller primary beam of these data.

Figure~\ref{fig2}c gives a detailed comparison of the \ammonia\/
(1,1), (3,3), and (6,6) integrated intensities. Significant (3,3)
emission is detected toward both the NE and SW core regions, though
the strongest emission is compact and located East of the EGO, toward
the SE terminus of the 44~GHz maser arc. Interestingly, this region
shows only weak para-\ammonia\/ emission suggesting a non-thermal
origin for the (3,3) emission, though it appears to be connected to
CM2 through a weaker ridge of warm thermal \ammonia\/ emission. The
(6,6) transition is also detected at the (3,3) peak, as well as toward
the candidate HC~\HII\/ region CM2.

We also detect strong, spatially-unresolved emission in the
\methanol\ $5_2-5_1$ and $8_2-8_1$ lines at two distinct positions
(Fig.~\ref{fig3}).  Using an upper limit to the fitted size of the
emission ($1\farcs5\times 0\farcs6$), their peak flux densities
($187\pm 2$ \mjb\/ and $39\pm 2$ \mjb\/) correspond to brightness
temperatures ($T_b$) exceeding 400 and 85 K for the stronger and
weaker source, respectively.  These high $T_b$ compared to the energy
levels above ground (57 and 105~K) strongly suggest that these
transitions are masing \citep[25 GHz \methanol\/ masers have also been
  observed for example in
  IRAS16547-4247,][]{Voronkov06}. Interestingly, the transition of
peak maser intensity switches from $5_2-5_1$ for the stronger maser
spot, to $8_2-8_1$ for the weaker maser spot.  The positions and peak
velocities are (J2000) 18$^{\rm h}$54$^{\rm m}$01.043$^{\rm s}$ ($\pm
0.002$s), $+$02$\arcdeg$01$\arcmin$16$\farcs$86 ($\pm 0\farcs 02$) at
52.8 \kms\/, and 18$^{\rm h}$54$^{\rm m}$00.562$^{\rm s}$
($\pm0.002$s), $+$02$\arcdeg$01$\arcmin$17$\farcs$67 ($\pm 0\farcs
02$) at 55.2 \kms\/ for the strong and weak masers, respectively (see
Fig.~\ref{fig2}a,b,c).

No emission greater than 5$\sigma$ was detected for the SO$_2$,
DC$_3$N, HC$_5$N, and HC$_9$N transitions.  The H63$\alpha$ and
H64$\alpha$ recombination lines were detected toward the CM1 UC~\HII\/
region at a peak velocity of 55.9 \kms\/. Following the method of
\citet{Garay86}, we use the line-to-continuum intensity ratio of 0.46,
the fitted FWHM linewidth of 17.6 \kms, and the diameter of CM1
determined from the EVLA 1.3~cm continuum ($1\farcs3$) to derive an
electron temperature of 7900~K and density of $1.3 \times 10^4$ \cc.



\subsection{\ammonia\/ Temperatures}


Over the last few decades considerable progress has been made in
understanding the excitation of the \ammonia\/ molecule under
astrophysical conditions \citep[see e.g.][]{Maret09}. For example in
cold regions, analysis of the (1,1) and (2,2) transitions has evolved
toward simultaneous least squares-fitting of the line profiles --
freeing the analysis from previous failures wherever the intensity of
(2,2) rivals (1,1) \citep[e.g.][]{Rosolowsky08}. However, this type of
analysis typically assumes that the kinetic temperature is much less
than the energy gap between levels (41.5 K) such that only the (1,1)
and (2,2) rotational levels are populated. Due to these assumptions,
this technique becomes less accurate above temperatures of about
30~K. Generic molecular modeling packages exist that can be used to
model \ammonia\/ emission from higher lying transitions using either
radiative transfer with collisional excitation, or assuming
LTE. However, no package currently combines separate ortho and para
collision rates (with realistic collision partners) with the detailed
hyperfine structure of \ammonia\/ and a partition function that
includes all relevant states that may be excited in warm gas (though
we can expect this situation to improve in the near future).

\begin{figure}
\epsscale{1.0}
\plotone{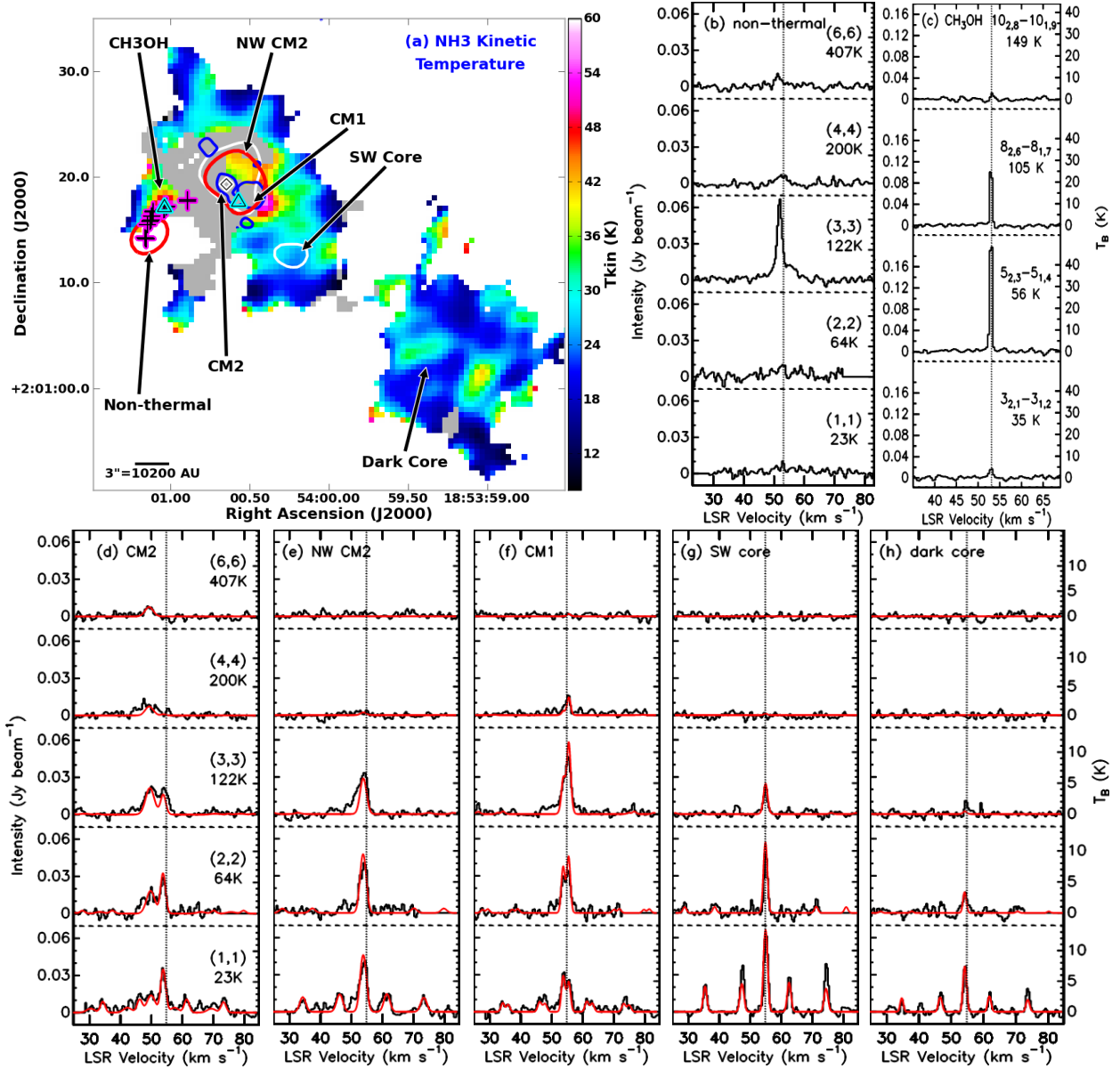}
\caption{(a) Colorscale of the \ammonia\/ kinetic temperature derived
  from single Gaussian two-level fits; grey areas indicate where the 
  assumptions for this
  technique are invalid. The blue, red, and white contours show the
  lowest 3.6~cm contour from Fig.~\ref{fig2}, highest (3,3) contour
  from Fig.~\ref{fig2}c, and highest (1,1) contour from
  Fig.~\ref{fig1}a, respectively. The locations of the profiles shown
  in (b), for the non-thermal \ammonia\/ emission (18$^{\rm
    h}$54$^{\rm m}$01.119$^{\rm s}$,
  $+$02$\arcdeg$01$\arcmin$14$\farcs$08) and (c) the stronger 25~GHz
  \methanol\/ maser (18$^{\rm h}$54$^{\rm m}$01.076$^{\rm s}$,
  $+$02$\arcdeg$01$\arcmin$16$\farcs$47) are indicated along with
  other positions fitted with method (2) and presented in
  Table~\ref{fits}. On (b) and (c), the vertical dotted line marks the
  LSR velocity of +53.1 \kms. (d)-(h) Observed \ammonia\ spectra (black) 
  overlaid with fitted profiles (red) using full LTE modeling in 
  CASSIS (see Table~\ref{fits}).
 \label{fig3}}
\end{figure}

For the current preliminary analysis we have pursued the following
course: (1) in the colder regions of G35.03 we have fit simultaneously
the full hyperfine structure of the (1,1) and (2,2) lines using a
non-linear, least squares Gaussian fitting routine given a single
common line-of sight-velocity, line width, excitation, and kinetic
temperatures in IDL \citep{Friesen09}; (2) In the warmer, and
coincidentally more kinematically complex regions we have used the LTE
method in the CASSIS\footnote{CASSIS has been developed by
  CESR-UPS/CNRS (http://cassis.cesr.fr).} package to model all six
observed \ammonia\/ transitions, including multiple velocity
components where necessary. To compare methods, we also fit the cooler
regions with CASSIS. Figure~\ref{fig3}a shows the results from
technique (1), with grey areas indicating where this method's
assumptions are invalid.  The fitted parameters for both techniques
toward several representative positions are given in Table~\ref{fits}.
The corresponding spectral line fits for these positions using
technique (2) are shown in Figure~\ref{fig3}d-h.

As shown in Table~\ref{fits}, the temperatures in the
kinematically-simple Dark and SW core regions are in the 20-30 K
range, with the latter being somewhat warmer on average. Both methods
do a reasonable job here, with method (2) predicting about twice the
total column density of (1), likely due to its more complete partition
function. Toward the center of the NE core and along the 44~GHz maser
arc, technique (1) breaks down for two reasons: (i) the gas is warm
($\gtrsim 30$ K) throughout this region (or even non-thermal;
Fig.~\ref{fig3}b) and (ii) the lines become more kinematically complex
with at least two distinct velocity components separated by about 1.7
\kms\/, or in the case of CM2 what appears to be a significantly
blueshifted hot (220 K) outflow component. Towards CM1, the redder
velocity component shows temperatures as high as $\sim 70$ K.

\subsection{Dust Mass}

The average column densities and masses of the NE and SW cores based
on the methodology described in \citet[][in Eqs.~4 and 5]{Brogan09}
using the 1.3~mm SMA data are shown in Table~\ref{fits}. The
integrated 1.3~mm flux densities of the NE and SW cores are 831 and
170 mJy with estimated sizes of $11.0\arcsec$ and $6.8\arcsec$,
respectively. For the NE core, we have subtracted 25 mJy to account
for the 3.6~cm flux of CM1 and CM2 (see Fig.\ref{fig2}) assuming they
are due entirely to free-free emission and using the 1.3~cm to 3.6~cm
spectral indices derived by \citet{Cyganowski11}.  For the range of
temperatures derived from the \ammonia\/ fits described in \S3.1, the
NE and SW cores have dust-based gas masses in the 50 - 132 M$_{\sun}$
and 22 - 45 M$_{\sun}$ range, respectively, and average column
densities of $(3-7)\times 10^{22}$ \ct\/. For comparison,
\citet{Hill05} report a SEST SIMBA 1.2~mm based mass for the whole
G35.03 region of 390 M$_{\sun}$ assuming a dust temperature of 20~K
and a total size of $55\arcsec$ (includes the ``Dark core'' in
addition to the NE and SW cores). The Bolocam Galactic Plane Survey
(BGPS) catalog\footnote{BGPS:
  http://irsa.ipac.caltech.edu/data/BOLOCAM\_GPS/} reports a 1.1~mm
flux of 1380 and 3030 mJy for G35.03 within $40\arcsec$ and
$80\arcsec$ apertures, respectively \citep{Rosolowsky10}. Assuming
$T_{dust}=30$ K for $r< 20\arcsec$ and $T_{dust}=20$~K for
$20\arcsec<r< 40\arcsec$, we find a total mass for the region of 350
M$_{\sun}$ (550 M$_{\sun}$ if we multiply by the BGPS ''correction
factor'' of 1.5), in reasonable agreement with \citet{Hill05} and our
SMA results, accounting for both the smaller primary beam and spatial
filtering of the interferometer.


\section{Discussion} 

\subsection{Protocluster Nature of G35.03+0.35}

Our high angular resolution observations reveal the presence of at
least three MYSOs (possibly four if CM3 is included) within 20,000 AU
of one another in the NE core (CM1, CM2, and NWCM2).  This
configuration suggests a Trapezium-like protocluster of massive stars
like those identified in other similar regions
\citep[e.g.][]{Hunter06,Rodon08}.  Component CM1 is a modest UC
\HII\ region associated with warm \ammonia\ and hosts two velocity
components which may indicate further unresolved structure.  CM2
appears to be a HC\HII\ region or wind/jet source exciting OH, H$_2$O,
and Class II \methanol\ masers, and a hot (220~K) blueshifted outflow.
The millimeter continuum source NW of CM2 exhibits the peak dust
emission, no \cmc\/ emission, but the broadest Gaussian (non-outflow)
\ammonia\ profiles.  Such diversity among the continuum properties and
in the molecular gas temperatures toward these different objects is an
increasingly familiar pattern seen in protoclusters
\citep{Zhang07,Brogan07}.  Similar to NGC6334I(N) \citep{Brogan09},
the systemic velocities of these objects differ by up to 1.7 \kms,
providing a measure of the cluster dynamics.  However, detailed study
of the kinematics requires higher angular resolution followup
observations, in particular to spatially resolve the multiple velocity
components toward the NE core objects.

\subsection{New Masers in G35.03+0.35}

As Figs.~\ref{fig2}c, and~\ref{fig3}b demonstrate, the (3,3) and (6,6)
\ammonia\/ emission toward the terminus of the 44~GHz \methanol\ maser
arc shows significant departure from LTE. Unfortunately, our current
angular resolution precludes the measurement of the high brightness
temperatures required for absolute confirmation of maser
emission. \citet{Walmsley83} first predicted that the (3,3) transition
could undergo weak population inversion for densities of $n\sim
10^{4-5}$ \cc\/ and temperatures of $\sim 50$ K -- physical conditions
common in massive molecular outflows and similar to the conditions
required to excite 44 GHz \methanol\/ masers \citep{Voronkov05}. We
speculate that the non-thermal \ammonia\/ emission traces a bow
shock from an outflow originating from CM2. Indeed, (3,3) maser emission
has been detected previously in several regions of massive star
formation mostly associated with strong outflows \citep[e.g. DR 21(OH),
  NGC~6334I, IRAS20126+4104, G5.89-0.39][]{Mangum94,Kraemer95,Zhang99,Hunter08}.

The stronger 25~GHz \methanol\/ maser spot (Fig.~\ref{fig3}c) is
coincident in position and velocity with one of the weaker 44~GHz
Class I masers from \citet{Cyganowski09}, and is within $1.3\arcsec$
of the strongest 44~GHz maser.  The weaker 25~GHz maser spot is
located towards the SE edge of the CM1 UCHII region. Class I
\methanol\/ masers are thought to be excited by collisions in shocks,
with the shock liberating methanol from dust grains and thus providing
the necessary high column density. \citet{Sobolev05} suggest that
25~GHz Class I masers require densities of $\sim 10^{5-7}$ \cc\/ and
temperatures of 75-100 K, hotter and denser than that required for
44~GHz class I masers. As demonstrated by \citet{Voronkov06} detailed
modeling of the ratio of brightness temperatures among the 25~GHz
masing transitions can help to pinpoint the physical conditions. We
hope to carry out such analysis for the ensemble of 25~GHz masers
observed across our survey in the future.

\section{Summary}

We present the first results from a 1.3~cm survey of MYSOs using the
EVLA, focusing on the EGO source G35.03+0.35.  The new EVLA correlator
allows us to simultaneously observe the \ammonia\/ (1,1) to (6,6)
transitions, four \methanol\/ transions, two H recombination lines,
and continuum. We find three major regions of dense \ammonia\/ gas,
two (the NE and SW cores) being coincident with the EGO, and the third
located in the adjacent IRDC. Using complementary SMA 1.3~mm and VLA
3.6~cm continuum data, along with temperatures derived from the
\ammonia\/ emission, we find warm gas temperatures in the range 20-220
K, core masses in the 20-130 M$_{\sun}$ range, and average H$_2$
column densities of several $\times 10^{22}$ \ct\/. Together these
data reveal a massive protocluster in an early stage of formation with
members representing different phases of MYSO evolution
concurrently. This work highlights the potential diagnostic power that
the full survey will provide toward understanding the formation of
massive stars and protoclusters.

\bigskip

\acknowledgments

Based on analysis using the CDMS, JPL, and Splatalogue spectroscopic
databases, NASA's Astrophysics Data System Bibliographic Services, and
the SIMBAD database.



\begin{thebibliography}{}

\bibitem[Argon et al.(2000)]{Argon00} Argon, A.~L., Reid,
M.~J., \& Menten, K.~M.\ 2000, \apjs, 129, 159

\bibitem[Brogan et al.(2007)]{Brogan07} Brogan, C.~L., Chandler,
C.~J., Hunter, T.~R., Shirley, Y.~L.,
\& Sarma, A.~P.\ 2007, \apjl, 660, L133

\bibitem[Brogan et al.(2009)]{Brogan09} Brogan, C.~L., Hunter, 
T.~R., Cyganowski, C.~J., Indebetouw, R., Beuther, H., Menten, K.~M., 
\& Thorwirth, S.\ 2009, \apj, 707, 1 

\bibitem[\protect\astroncite{{Cyganowski} et~al.}{2011}]{Cyganowski11}
{Cyganowski}, C.~J., {Brogan}, C.~L., {Hunter}, T.~R., and {Churchwell}, E.:
  2011,
\newblock {\em \apj} 00

\bibitem[\protect\astroncite{{Cyganowski} et~al.}{2009}]{Cyganowski09}
{Cyganowski}, C.~J., {Brogan}, C.~L., {Hunter}, T.~R., and {Churchwell}, E.:
  2009,
\newblock {\em \apj} {\bf 702}, 1615

\bibitem[\protect\astroncite{{Cyganowski} et~al.}{2008}]{Cyganowski08}
{Cyganowski}, C.~J., et al.\ 2008,
\newblock {\em \aj} {\bf 136}, 2391


\bibitem[\protect\astroncite{{Forster} and {Caswell}}{1999}]{Forster99}
{Forster}, J.~R. and {Caswell}, J.~L.: 1999,
\newblock {\em \aaps} {\bf 137}, 43

\bibitem[Friesen et al.(2009)]{Friesen09} Friesen, R.~K., Di 
Francesco, J., Shirley, Y.~L., \& Myers, P.~C.\ 2009, \apj, 697, 1457 

\bibitem[\protect\astroncite{{Garay} et~al.}{1986}]{Garay86}
{Garay}, G., {Rodriguez}, L.~F., and {van Gorkom}, J.~H.: 1986,
\newblock {\em \apj} {\bf 309}, 553

\bibitem[Hill et al.(2005)]{Hill05} Hill, T., Burton, M.~G., 
Minier, V., Thompson, M.~A., Walsh, A.~J., Hunt-Cunningham, M., 
\& Garay, G.\ 2005, \mnras, 363, 405 

\bibitem[\protect\astroncite{{Hunter} et~al.}{2008}]{Hunter08}
{Hunter}, T.~R., {Brogan}, C.~L., {Indebetouw}, R., and {Cyganowski}, C.~J.:
  2008,
\newblock {\em \apj} {\bf 680}, 1271

\bibitem[Hunter et al.(2006)]{Hunter06} Hunter, T.~R., Brogan,
C.~L., Megeath, S.~T., Menten, K.~M., Beuther, H.,
\& Thorwirth, S.\ 2006, \apj, 649, 888 


\bibitem[\protect\astroncite{{Kraemer} and {Jackson}}{1995}]{Kraemer95}
{Kraemer}, K.~E. and {Jackson}, J.~M.: 1995,
\newblock {\em \apjl} {\bf 439}, L9

\bibitem[\protect\astroncite{{Kurtz} et~al.}{1994}]{Kurtz94}
{Kurtz}, S., {Churchwell}, E., and {Wood}, D.~O.~S.: 1994,
\newblock {\em \apjs} {\bf 91}, 659

\bibitem[\protect\astroncite{{Mangum} and {Wootten}}{1994}]{Mangum94}
{Mangum}, J.~G. and {Wootten}, A.: 1994,
\newblock {\em \apjl} {\bf 428}, L33

\bibitem[Maret et al.(2009)]{Maret09} Maret, S., Faure, A., 
Scifoni, E., \& Wiesenfeld, L.\ 2009, \mnras, 399, 425 

\bibitem[Perley et al.(2011)]{Perley11} 
Perley, R.A., Chandler, C.J., Butler, B.J., Wrobel, J.M. 2011, \apjl, in press 

\bibitem[\protect\astroncite{{Rathborne} et~al.}{2006}]{Rathborne06}
{Rathborne}, J.~M., {Jackson}, J.~M., and {Simon}, R.: 2006,
\newblock {\em \apj} {\bf 641}, 389

\bibitem[Rod{\'o}n et
al.(2008)]{Rodon08} Rod{\'o}n, J.~A., Beuther, H., Megeath, S.~T., \& van der T\
ak, F.~F.~S.\ 2008, \aap, 490, 213 

\bibitem[Rosolowsky et al.(2008)]{Rosolowsky08} Rosolowsky, E.~W., 
Pineda, J.~E., Foster, J.~B., Borkin, M.~A., Kauffmann, J., Caselli, P., 
Myers, P.~C., \& Goodman, A.~A.\ 2008, \apjs, 175, 509 

\bibitem[\protect\astroncite{{Rosolowsky} et~al.}{2010}]{Rosolowsky10}
{Rosolowsky}, E., et al.\ 2010,
\newblock {\em \apjs} {\bf 188}, 123

\bibitem[\protect\astroncite{{Sobolev} et~al.}{2005}]{Sobolev05}
{Sobolev}, A.~M., {Ostrovskii}, A.~B., {Kirsanova}, M.~S., {Shelemei}, O.~V.,
  {Voronkov}, M.~A., and {Malyshev}, A.~V.: 2005,
\newblock in {R.~Cesaroni, M.~Felli, E.~Churchwell, \& M.~Walmsley} (ed.), {\em
  Massive Star Birth: A Crossroads of Astrophysics}, Vol. 227 of {\em IAU
  Symposium}, pp 174--179


\bibitem[Voronkov et al.(2005)]{Voronkov05} 
Voronkov, M., Sobolev, A., Ellingsen, S., Ostrovskii, A., \& Alakoz, A.\ 2005, \apss, 295, 217 

\bibitem[\protect\astroncite{{Voronkov} et~al.}{2006}]{Voronkov06}
{Voronkov}, M.~A., {Brooks}, K.~J., {Sobolev}, A.~M., {Ellingsen}, S.~P.,
  {Ostrovskii}, A.~B., and {Caswell}, J.~L.: 2006,
\newblock {\em \mnras} {\bf 373}, 411

\bibitem[\protect\astroncite{{Walmsley} and {Ungerechts}}{1983}]{Walmsley83}
{Walmsley}, C.~M. and {Ungerechts}, H.: 1983,
\newblock {\em \aap} {\bf 122}, 164

\bibitem[Zhang et al.(2007)a]{Zhang07} Zhang, Q., et al.\ 2007, \apj,
  658, 1152

\bibitem[Zhang et al.(1999)]{Zhang99} Zhang, Q., Hunter, T.~R., 
Sridharan, T.~K., \& Cesaroni, R.\ 1999, \apjl, 527, L117 

\bibitem[\protect\astroncite{{Zinnecker} and {Yorke}}{2007}]{Zinnecker07}
{Zinnecker}, H. and {Yorke}, H.~W.: 2007,
\newblock {\em \araa} {\bf 45}, 481

\end{thebibliography}

\begin{deluxetable}{ll}
\tablewidth{0pc}
\tablecaption{Observing parameters\label{smaobs}}  
\tablecolumns{2}
\tablehead{\colhead{Parameter} & \colhead{Value}}
\startdata
\cutinhead{EVLA 1.3~cm ($\sim 24$ GHz) AB1346 observations}
Observing date (duration) & 07 Sep 2010 (3 hours)\\
Configuration & D\\
Primary beamsize & $2\arcmin$ \\
Bandwidth & $16\times$ 8 MHz, single polarization \\ 
Baseband 0 subbands (MHz),\tablenotemark{a} lines
                           & 23692.78, \ammonia\ (1,1)\\
                           & 23724.78, \ammonia\ (2,2)\\
                           & 23828.78, OH $^2\Pi_{9/2}$ ($5^--5^+$)\\  
                           & 23868.78, \ammonia\ (3,3)\\ 
                           & 24084.78, SO$_2$ ($8_{2,6}-9_{1,9}$)\\
                           & 24140.78, \ammonia\ (4,4)\\
                           & 24508.78, H64$\alpha$\\
                           & 24532.78, \ammonia\ (5,5)\\
Baseband 1 subbands (MHz),\tablenotemark{a} lines
                           & 24927.88, \methanol\ ($3_{2,1}-3_{1,2}$)\\
                           & 24959.88, \methanol\ ($5_{2,3}-5_{1,4}$)\\
                           & 25023.88, NH$_2$D ($4_{1,4}-4_{0,4}$)\\ 
                           & 25055.88, \ammonia\ (6,6)\\
                           & 25295.88, \methanol\ ($8_{2,6}-8_{1,7}$)\\
                           & 25327.88, DC$_3$N (3-2)\\
                           & 25687.88, H63$\alpha$\\
                           & 25979.88, \methanol\ ($10_{2,8}-10_{1,9}$)\\
Velocity resolution & 0.4 \kms\\
Angular resolution\tablenotemark{b} & $3\farcs7\times 3\farcs$0 (P.A.=$-50\arcdeg$)\\
Spectral line rms noise\tablenotemark{c}  & 3 \mjb\/ channel$^{-1}$ \\
Gain calibrator &  J1815+0035 \\
Bandpass and Flux Calibrator\tablenotemark{d} & J1924-2914 (17.1 Jy assumed)\\
\\
\\
\\
\cutinhead{SMA 1.3~mm ($\sim 225$ GHz) observations}
Observing date (duration) & 24 Jun 2008 (11 hours)\\
Configuration & compact-north\\
Primary beamsize & $52\arcsec$ \\
Bandwidth & $2\times$ 2 GHz, single polarization\\
Velocity resolution & 1.1 \kms \\
Angular resolution$^{b}$ & $3\farcs2\times 1\farcs8$ (P.A.=$+70\arcdeg$)\\
Continuum rms noise & 3 \mjb\/  \\
Gain Calibrators &  J1733-130 \& J1751+096 \\
Bandpass Calibrator & 3C454.3 \\
Flux Calibrator & Uranus 
\enddata
\tablenotetext{a}{These are the rest frame subband center frequencies;
  because the subbands were required to be on an 8 MHz grid, the lines
  are offset from the centers.  Line rest frequencies were obtained
  from http://splatalogue.net. }
\tablenotetext{b}{Briggs weighting of 0.5.}
\tablenotetext{c}{Due to its inadvertent proximity to a subband filter
  edge, the \ammonia\ (2,2) subband is a factor
  of two noisier than the others.  } 
\tablenotetext{d}{3C~286 observation failed, used bootstrapped flux 
  density of J1924-2914 from previous track.}
\end{deluxetable}

\begin{deluxetable}{lccccccccccc}
\tabletypesize{\scriptsize}
\rotate
\tablecaption{Fitted Parameters\label{fits}}  
\tablewidth{0pt}
\tablehead{
\colhead{ } & \multicolumn{4}{c}{Single Gaussian 2-level \ammonia\/ Fits} & \multicolumn{4}{c}{Full Multitransition CASSIS \ammonia\/ Fits} 
& \multicolumn{3}{c}{Estimates from 1.3~mm Continuum} \\
\colhead{Source$^a$}  & \colhead{$V_{LSR}$} &  \colhead{$\Delta V_{FWHM}$} & \colhead{$T_k$} & \colhead{$N_{NH_3}$} & 
\colhead{$V_{LSR}$} & \colhead{$\Delta V_{FWHM}$} & \colhead{$T_k$} & \colhead{$N_{NH_3}$} & 
\colhead{$T_{dust}$} & \colhead{$ M_{gas}$} & \colhead{$N_{H_2}$} \\
\colhead{ }       &   \colhead{(\kms\/)} &  \colhead{(\kms\/)} &  \colhead{(K)}  &  \colhead{($\times 10^{15}$ \ct\/)} &  
\colhead{(\kms\/)} &  \colhead{(\kms\/)} &   \colhead{(K)}  & \colhead{($\times 10^{15}$\ct\/)} & 
\colhead{(K)} &  \colhead{(M$_{\sun}$)} & \colhead{($\times 10^{22}$\ct\/)} }
\startdata
NE Core    & \nodata  & \nodata  & \nodata  & \nodata  & \nodata  & \nodata  & \nodata  & \nodata  & 70 - 30    & 50 - 132   & 3 - 7 \\
~~~~CM1-v1 & \nodata  & \nodata  & \nodata  & \nodata  & 53.7  & 1.3  & 40  & 2.0  & \nodata  & \nodata  & \nodata \\
~~~~CM1-v2 & \nodata  & \nodata  & \nodata  & \nodata  & 55.4  & 1.3  & 70  & 3.2  & \nodata  & \nodata  & \nodata \\
~~~~CM2-v1 & \nodata  & \nodata  & \nodata  & \nodata  & 53.8  & 1.8  & 30  & 3.5  & \nodata  & \nodata  & \nodata \\
~~~~CM2-v2 & \nodata  & \nodata  & \nodata  & \nodata  & 50.0  & 3.0  & 35  & 2.0  & \nodata  & \nodata  & \nodata \\
~~~~CM2-v3 & \nodata  & \nodata  & \nodata  & \nodata  & 49.2  & 3.0  & 220 & 2.0  & \nodata  & \nodata  & \nodata \\
~~~~NW CM2 & 53.3  & 2.8  & 41  & 2.1  & 53.8  & 2.2  & 33  & 6.0  & \nodata  & \nodata  & \nodata \\
SW Core    & 54.7  & 1.1  & 25  & 3.0  & 54.9  & 1.2  & 27  & 5.5  & 35 - 20    & 22 - 45    & 3 - 6  \\
Dark Core  & 53.9  & 1.0  & 20  & 1.2  & 54.2  & 1.1  & 18  & 2.6  & \nodata  & \nodata  &  \nodata 
\enddata
\tablenotetext{a}{Positions for these fits are (J2000):
CM1:  18$^{\rm h}$54$^{\rm m}$00.488$^{\rm s}$, $+$02$\arcdeg$01$\arcmin$18$\farcs$21; 
CM2: 18$^{\rm h}$54$^{\rm m}$00.650$^{\rm s}$, $+$02$\arcdeg$01$\arcmin$19$\farcs$05; 
NW~CM2: 18$^{\rm h}$54$^{\rm m}$00.561$^{\rm s}$, $+$02$\arcdeg$01$\arcmin$21$\farcs$72; 
SW core: 18$^{\rm h}$54$^{\rm m}$00.164$^{\rm s}$, $+$02$\arcdeg$01$\arcmin$11$\farcs$78; 
Dark Core: 18$^{\rm h}$53$^{\rm m}$59.396$^{\rm s}$,$+$02$\arcdeg$01$\arcmin$01$\farcs$47.}
\end{deluxetable}

\end{document}